\documentclass[preprint,18pt,GungsuhChe]{revtex4-1}
\usepackage{amsmath,natbib,graphicx,subfigure,amssymb,graphics,amsmath,mathrsfs,CJK,color}
\usepackage{multirow,fancyhdr,color,bm,tabularx,psfrag,dcolumn}
\usepackage[colorlinks=true,linkcolor=blue,urlcolor=blue,citecolor=blue]{hyperref}
\usepackage[left]{lineno}
\usepackage{tikz,xcolor,hyperref}
\definecolor{lime}{HTML}{A6CE39}
\DeclareRobustCommand{\orcidicon}{%
    \begin{tikzpicture}
    \draw[lime, fill=lime] (0,0)
    circle [radius=0.16]
    node[white] {{\fontfamily{qag}\selectfont \tiny ID}};\draw[white, fill=white] (-0.0625,0.095)
    circle [radius=0.007];
    \end{tikzpicture}
    \hspace{-2mm}}
\foreach \x in {A, ..., Z}
{\expandafter\xdef\csname orcid\x\endcsname{\noexpand\href{https://orcid.org/\csname orcidauthor\x\endcsname}{\noexpand\orcidicon}}}

\begin{document}
\title{\LARGE Photovoltaic properties enhanced by the tunneling effect in a coupled quantum dot photocell}


\author{Sheng-Qiang Zhong }
\affiliation{Department of Physics, Faculty of Science, Kunming University of Science and Technology, Kunming, 650500, PR China}

\author{Shun-Cai Zhao\orcidA{}}
\email[Corresponding author: ]{zhaosc@kmust.edu.cn.}
\affiliation{Department of Physics, Faculty of Science, Kunming University of Science and Technology, Kunming, 650500, PR China}

\author{Sheng-Nan Zhu }
\affiliation{Department of Physics, Faculty of Science, Kunming University of Science and Technology, Kunming, 650500, PR China}
\date{\today}
\begin{abstract}
Double quantum dots (DQDs) have emerged as versatile and efficient absorbing light devices owing to their more multiple adjusting parameters than the single QD's. Using the system-reservoir theory, tunneling effect on the quantum photovoltaic properties is evaluated detailedly in a DQDs photocell with different energy mismatches under different temperatures. Repopulation of photo-generated carriers in two excited states is generated by electron tunneling effect between two coupled QD pairs, which brings out more efficient quantum photovoltaic yields evaluated by the short-circuit current, open-circuit voltage and output power through the current-voltage and power-voltage characteristics. Further discussion reveals that the quantum photovoltaic properties are enhanced by the ambient environment temperatures but weakened by the energy mismatches. However, the weakening effect caused by the energy mismatches can be greatly reduced by tunneling effect. Insights into tunneling effect between two adjacent QD not only facilitate a better understanding of the fundamentals in carriers transport, but also provide novel strategies for efficient artificial assembled QD arrays photocell inspired by this proposed DQDs photocell model.
\begin{description}
\item[PACS: ]{42.50.Gy  }
\item[Keywords:]{Photovoltaic properties, tunneling effect, coupled quantum dot photocell }
\end{description}
\end{abstract}

\maketitle

\section{Introduction }

Designing an artificial system to mimic nature efficiently converting solar energy to other energy, such as electricity, can meet the increasing energy demand in current society. Quantum dots (QDs) are identified as an efficient artificial system\cite{2018Photocatalysis,2017General,Xiang2019Quantum,2019Regio} to capture light and convert the energy to other useful forms, owing to their quantum confinement effects\cite{2013Third}. Hence, their optical and electrical properties can be tuned by the physical parameters, such as its size and shape. Not only that, their large absorption coefficients\cite{2010Comparing,2018Optical}, in conjunction with broad and tunable absorption spectra, make QDs excellent solar absorbers. When QDs are combined with another semiconductor or a metal contact, a heterostructure is formed\cite{2010Stability,2019High}. Therefore, the photo-generated carriers in the QD layers can be separated, which performs as a photovoltaic device. In this way, QDs are used to convert optical energy into electrical energy.

A lots of work on efficient photoelectric conversion efficiency of QD photovoltaic cells has been carried out\cite{Zhao2019,2013Quantum,Mcdaniel2013Correction,2013Sensitized,2002Quantum,2016Iodide,2019Efficient,2020High,Wu2019Quantum,2020Radiative}. Recently, QDs have been believed to be an alternative efficient candidate of solar cell via a multilevel structure\cite{Zhao2019,2013Quantum,2020High}, due to their ability to enhance photon harvesting. The absorption spectrum of QD can be tuned down to the infrared range because of its adjustable dot size\cite{2002Quantum}, which makes QD competitive in designing multi-junction solar cells. In addition, the carrier density is increased near the QD surface owing to the strong quantum confinement effect\cite{Mcdaniel2013Correction,2013Sensitized,2016Iodide,2019Efficient,2010Comparing}, which facilitates the interfacial charge/energy transfer between QDs and nearby molecular acceptors. Another strategy is adopted to separate electrons and holes for reductive process avoiding oxidation\cite{Wu2019Quantum} integrating the QDs into a photo-electrochemical cell. In addition to these, multiple exciton generation is demonstrated to be more efficient in QDs\cite{2010Comparing,2011Peak} than bulk semiconductors in single-junction solar cells.

However, none of these studies focus on the improvement of the quantum yields via the tunneling effect between the double QDs from the perspective of quantum optics. Thus, in this paper, we aim to the tunneling effect on the current-voltage and power-voltage characteristics in a DQDs photocell. And almost the same populated photo-generated carriers is revealed by the robust tunneling effect in the excited states, which brings out the efficient quantum yields in this DQDs photocell. It may provide some artificial strategies for efficient assembled QD arrays photocell inspired by the DQDs photocell.

\section{Theoretical model for the DQDs photocell}

\begin{figure}[htp]
\includegraphics[width=0.30\columnwidth] {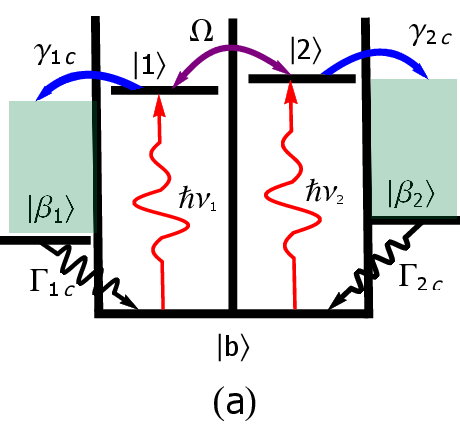}~~~~~~~~~~~\includegraphics[width=0.38\columnwidth] {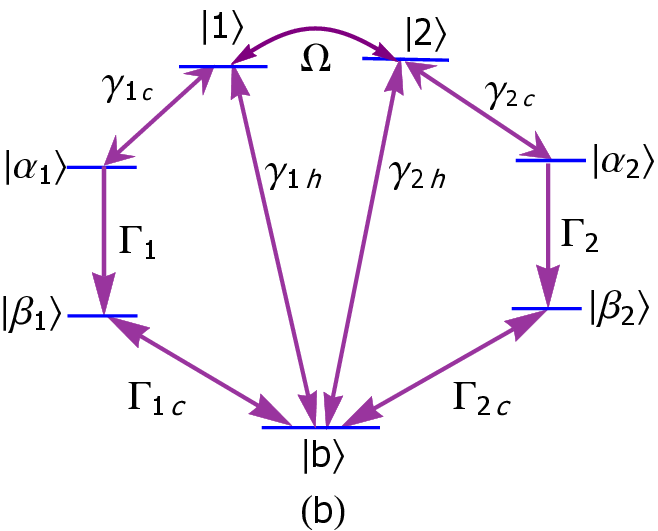}
\caption{(Color online)(a) Electronic states of the DQDs are presented in their eigenstate basis via the solid lines, and they connect with each other via tunneling coefficient \(\Omega\). Two photons, \(\hbar\nu_{i=1,2}\) are absorbed by this DQDs, respectively. The excited photocurrents release out via the rates \(\gamma_{ic(i=1,2)}\) to the external terminals. The left/right lead, \(|\beta_{1,2}\rangle\) with rate \(\Gamma_{ic(i=1,2)}\) relaxes to the ground state, \(|b\rangle\). (b) Corresponding energy-level diagram for this DQDs photocell model. The absorbed photons drive transitions through \(|1\rangle\) \(\leftrightarrow\) \(|b\rangle\) and \(|2\rangle\) \(\leftrightarrow\) \(|b\rangle\). Transitions \(|1\rangle\) \(\leftrightarrow\) \(|\alpha_{1}\rangle\) and \(|2\rangle\) \(\leftrightarrow\) \(|\alpha_{2}\rangle\), \(|\beta_{1}\rangle\) \(\leftrightarrow\) \(|b\rangle\) and \(|\beta_{2}\rangle\) \(\leftrightarrow\) \(|b\rangle\) are driven by the ambient thermal phonons. Two loads are connected to two levels \(|\alpha_{i}\rangle\) and \(|\beta_{i}\rangle_{(i=1,2)}\).}
\label{Fig.1}
\end{figure}

We consider a DQDs system with each dot coupling to its individual thermal reservoir in the solar environment, and the double dots connect with each other via the tunneling coefficient \(\Omega\), see Figs.\ref{Fig.1}(a). Then, the photovoltaic system Figs.\ref{Fig.1}(a) can be described by a multi-level quantum photocell model in Fig.\ref{Fig.1}(b). The transport of the carriers is initiated by the absorbed photons with energy higher than the band-gap energy. Thereupon, the exciton, i.e., the electron-hole pairs is generated. To be specific, we denote state $|b\rangle$ as the valence band (VB), and $|1\rangle$ ($|2\rangle$) as the conduction band (CB) in this DQDs photocell. Rates \(\gamma_{ih(i=1,2)}\) describe the carriers transporting speed, and the excited photocurrent releases out via the rates \(\gamma_{ic(i=1,2)}\) to the external terminals. The loads connected to $|\alpha_{i}\rangle$ \(\leftrightarrow\)$|\beta_{i}\rangle$ are yielding a decay of the levels $|\alpha_{i}\rangle$ \(\leftrightarrow\)$|\beta_{i}\rangle$ at a rate $\Gamma_{i}(i=1,2)$. We model the photovoltaic system by the Hamiltonian \(\hat{H}=\hat{H}_{DQD}+\hat{H}_{Lead}+\hat{H}_{Photon}+\hat{H}_{L-D}+\hat{H}_{P-D}\), where $\hat{H}_{DQD}$ denotes the Hamiltonian of the DQDs system, with an extra electron in the left or right dot (\(|1\rangle\) or \(|2\rangle\)),
\begin{equation}
\hat{H}_{DQD}=\frac{\varepsilon}{2}\hat{\sigma}_{z}+\Omega\hat{\sigma}_{x},
\end{equation}

\noindent with \(\varepsilon\) being the electrostatic energy mismatch\cite{2013Quantum} between the two QDs, and \(\Omega\) being the tunneling coefficient of an electron between the double dots. The Pauli operators are defined as $\hat{\sigma}_{z}$ = $\hat{d}_{L}^{\dag}$$\hat{d}_{L}$- $\hat{d}_{R}^ {\dag} \hat{d}_{R}$, $\hat{\sigma}_{x}$= $\hat{d}_{L}^ {\dag}$$\hat{d}_{R}$ + $\hat{d}_{R}^ {\dag}$$\hat{d}_{L}$\cite{Fujisawa932}, with $\hat{d}_{L}^ {\dag} ( \hat{d}_{L}$), $\hat{d}_{R}^ {\dag} ( \hat{d}_{R}$) respectively representing the creation (annihilation) operators in the left  or right quantum dot. In Eq.(2), $\hat{H}_{Lead}$ represents the Hamiltonian of the double electrodes, which is  modeled as collections of noninteracting electrons,
\begin{equation}
\hat{H}_{Lead}=\sum_{i=1,2}\sum_{k}\omega_{ik}\hat{C}_{ik}^ {\dag}\hat{C}_{ik},
\end{equation}
\noindent with $\hat{C}_{ik}^ {\dag}$($\hat{C}_{ik}$) depicting the creation (annihilation) operator for an electron with momentum k in the \emph{i}th electrode(\emph{i}=1,2). And the tunneling-coupling term between the dot $\nu$ and lead $\nu$($\nu$=1,2) is read as
\(\hat{H}_{L-D}=\sum_{\nu=1,2}\sum_{k}g_{\nu k}\hat{d}_{\nu k}^ {\dag}\hat{C}_{\nu k}+H.c\) with the corresponding coupling constant $g_{\nu k}$.
The ambient environment's  Hamiltonian can be written as $\hat{H}_{Photon}=\sum_{q}\omega_{q}\hat{b}_{q}^ {\dag}\hat{b}_{q}$, and the interacting Hamiltonian between the DQDs photovoltaic system and the ambient environment\cite{Wertnik2018Optimizing} with the corresponding coupling strength($\lambda_{q}$) are modeled as,
\begin{equation}
\widehat{H}_{P-D}=\hat{\sigma}_{x}\sum_{q}\lambda_{q}(\hat{b}_{q}^ {\dag}+\hat{b}_{q}).
\end{equation}

Therefore, the behavior of this QDQs photovoltaic system in the presence of relaxation in electron and photon degrees of freedom by employing the Born-Markov master equation for the reduced density matrix:
\begin{equation}
\frac{\partial}{\partial t}\hat{\rho}(t)=-i[\,\hat{H}_{s},\hat{\rho}\,]+\hat{{\cal L}}_{h}\hat{\rho}+\hat{{\mathcal{L}}}_{c_i}\hat{\rho}+\hat{{\mathcal{L}}}_{\Gamma c_i}\hat{\rho}+\hat{{\mathcal{L}}}_{rel}\hat{\rho},
\end{equation}
\noindent where the Hamiltonian \(\hat{H}_{s}\)=\(\hat{H}_{DQD}+\hat{H}_{Lead}+\hat{H}_{L-D}\) describe the DQDs photovoltaic system and coupling with the electrodes, the Hamiltonian \(\hat{H}_{Photon}+\hat{H}_{P-D}\) is considered as the ambient environment. The superoperator $\hat{\mathcal{L}}$ was decomposed into three parts for describing the effect of reservoirs and acceptor-donor re-combinations, including $\hat{{\cal L}}_{h}\hat{\rho}$, $\hat{{\mathcal{L}}}_{c_i}\hat{\rho},\hat{{\mathcal{L}}}_{\Gamma c_i}\hat{\rho},\hat{{\mathcal{L}}}_{rel}\hat{\rho}$. The dissipative processes between the DQDs photovoltaic system and the ambient solar environment can be denoted as

\begin{equation}
\hat{\mathcal{L}}_{h}\hat{\rho} = \frac{\gamma_{ih}}{2}[ (n_{ih}+1)\mathcal{D}[\hat{\sigma}_{bi}]\rho+n_{ih}\mathcal{D}[\hat{\sigma}_{bi}^{\dag}]\rho],
\end{equation}

\noindent where $\gamma_{ih}$ describes the dissipative effect, and the average photon numbers $n_{ih} = \frac{1}{exp[\frac{E_{i}b}{K_BT_s}]-1}$ are described as the effect of photon environment on the DQDs with the sun's temperature $T_s$. Moreover,the specific expression of the mark $\mathcal{D}$ acting on any operator $\hat{A}$ is $\mathcal{D}[\hat A]\hat\rho=2\hat A\hat\rho\hat A^{\dag}-\hat\rho\hat A^{\dag}\hat A-\hat A^{\dag}\hat A\hat\rho$. Superoperator $\hat{\mathcal{L}}_{c_1}\hat{\rho}$ corresponding to the quantum transition between the dot-lead coupling is described by
\begin{equation}
\hat{\mathcal{L}}_{c_i}\hat{\rho} = \frac{\gamma_{ic}}{2}[ (n_{ic}+1)\mathcal{D}[\hat{\sigma}_{\alpha_ii}]\rho+n_{ic}\mathcal{D}[\hat{\sigma}_{\alpha_ii}^{\dag}]\rho],
\end{equation}
\noindent where $\hat{\sigma}_{\alpha_ii}$ = $|\alpha_i\rangle\langle i|$ (i=1,2), and the corresponding average phonon numbers $n_{ic} = \frac{1}{exp[\frac{E_{i}\alpha_i}{K_B T_a}]-1}$, with the ambient temperature $T_a$. $\gamma_{ic}$ is the spontaneous decay rate from (to) the level $|i\rangle$ to (from) the level $|\alpha_i\rangle$. Similarly, the dissipative process between the left/right lead, \(|\beta_{1,2}\rangle\) and the ground state, \(|b\rangle\) can be written as,

\begin{equation}
\hat{\mathcal{L}}_{\Gamma c_i}\hat{\rho} = \frac{\Gamma_{c_i}}{2}[ (N_{ic}+1)\mathcal{D}[\hat{\sigma}_{\beta_{i}b}]\rho+N_{ic}\mathcal{D}[\hat{\sigma}_{\beta_{i}b}^{\dag}]\rho],
\end{equation}
In the same way,  $\hat{\sigma}_{\beta_{i}b}$ = $|\beta_i\rangle\langle b|$ with the average phonon numbers $N_{ic} = \frac{1}{exp[\frac{E_{b}\beta_i}{K_B T_a}]-1}$ the ambient temperature $T_a$,(i=1,2). Finally, the output of this DQDs photovoltaic system is described by a relaxation rates $\Gamma_i$ proportional to the electronic current, which decays from state $|\alpha_i\rangle$ to $|\beta_i\rangle$ as follows,
\begin{equation}
\hat{{\mathcal{L}}}_{rel}\hat{\rho} = \sum_{i=1,2}\frac{\Gamma_i}{2}(|\beta_i\rangle\langle \alpha_i|\hat\rho|\alpha_i\rangle\langle \beta_i|-|\alpha_i\rangle\langle \alpha_i|\hat\rho-\hat\rho|\alpha_i\rangle\langle \alpha_i|),
\end{equation}

\section{The quantum yields of the DQDs photovoltaic system }

Under the Weisskopf-Wigner approximation\cite{1974On}, we can obtain the dynamic equations of the corresponding reduced matrix elements as follows:
\begin{eqnarray}
&\dot{\rho}_{11}\!=\!&-i\Omega(\rho_{21}-\rho_{12})-\gamma_{1c}[(n_{1c}+1)\rho_{11}-n_{1c}\rho_{\alpha_{1}\alpha_{1}}]-\gamma_{1h}[(n_{1h}+1)\rho_{11}-n_{1h}\rho_{bb}]
-i\gamma_{1c}(\rho_{\alpha_{1}1}-\rho_{1\alpha_{1}}),\nonumber\\
&\dot{\rho}_{22}\!=\!&i\Omega(\rho_{21}-\rho_{12})-\gamma_{2c}[(n_{2c}+1)\rho_{22}-n_{2c}\rho_{\alpha_{2}\alpha_{2}}]-\gamma_{2h}[(n_{2h}+1)\rho_{22}-n_{2h}\rho_{bb}]\nonumber\\&&-i\gamma_{2c}(\rho_{\alpha_{2}2}-\rho_{2\alpha_{2}}),\nonumber\\
&\dot{\rho}_{\alpha_{1}\alpha_{1}}\!=\!&i\gamma_{1c}(\rho_{\alpha_{1}1}-\rho_{1\alpha_{1}})+\gamma_{1c}[(n_{1c}+1)\rho_{11}-n_{1c}\rho_{\alpha_{1}\alpha_{1}}]-\Gamma_{1}\rho_{\alpha_{1}\alpha_{1}},\nonumber\\
&\dot{\rho}_{\alpha_{2}\alpha_{2}}\!=\!&i\gamma_{2c}(\rho_{\alpha_{2}2}-\rho_{2\alpha_{2}})+\gamma_{2c}[(n_{2c}+1)\rho_{22}-n_{2c}\rho_{\alpha_{2}\alpha_{2}}]-\Gamma_{2}\rho_{\alpha_{2}\alpha_{2}},\nonumber\\
&\dot{\rho}_{\beta_{1}\beta_{1}}\!=\!&\Gamma_{1}\rho_{\alpha_{1}\alpha_{1}}-\Gamma_{1c}[(N_{1c}+1)\rho_{\beta_{1}\beta_{1}}-N_{1c}\rho_{bb}],\\
&\dot{\rho}_{\beta_{2}\beta_{2}}\!=\!&\Gamma_{2}\rho_{\alpha_{2}\alpha_{2}}-\Gamma_{2c}[(N_{2c}+1)\rho_{\beta_{2}\beta_{2}}-N_{2c}\rho_{bb}],\nonumber\\
&\dot{\rho}_{12}\!=\!&-i\epsilon\rho_{12}-i\Omega(\rho_{22}-\rho_{11})-\frac{\rho_{12}}{2}[\gamma_{1h}(n_{1h}+1)+\gamma_{2h}(n_{2h}+1)+\gamma_{1c}(n_{1c}+1)+\gamma_{2c}(n_{2c}+1)],\nonumber\\
&\dot{\rho}_{1\alpha_{1}}\!=\!&-i\gamma_{1c}(\rho_{\alpha_{1}\alpha_{1}}-\rho_{11})+i(\omega_1-\frac{\epsilon}{2}){\rho}_{1\alpha_{1}}-\frac{\rho_{1\alpha_1}}{2}[\gamma_{1h}(n_{1h}+1)+\gamma_{1c}(2n_{1c}+1)+\Gamma_1],\nonumber\\
&\dot{\rho}_{2\alpha_{2}}\!=\!&-i\gamma_{2c}(\rho_{\alpha_{2}\alpha_{2}}-\rho_{22})+i(\omega_2+\frac{\epsilon}{2}){\rho}_{2\alpha_{2}}-\frac{\rho_{2\alpha_2}}{2}[\gamma_{2h}(n_{2h}+1)+\gamma_{2c}(2n_{2c}+1)+\Gamma_2].\nonumber
\end{eqnarray}

And we focus on steady-state operation of this DQDs photocell model, voltage \(V\) across the load and the output effective electric current are expressed in terms of populations of the levels $|\alpha_i\rangle$ and $|\beta_i\rangle$, $eV$=$\sum\limits_{i=1,2}[\mu_{\alpha_i}-\mu_{\beta_i}+k_B T\ln(\frac{{\rho}_{\alpha_{i}\alpha_{i}}}{{\rho}_{\beta_i\beta_i}})]$ and $I$=$\sum\limits_{i=1,2}e\Gamma_i{\rho}_{\alpha_{i}\alpha_{i}}$\cite{2020Radiative,2020Inhibited}, respectively. Therefore, the current-voltage characteristic and photovoltaic power are crucial for analyzing the quantum photovoltaic effect in this DQDs photocell model, in which the short circuit current, open circuit voltage and output power can be explicitly identified. On account of the similar physical regime for the two different photovoltaic processes, we emphatically investigate the tunneling effect between the double QDs on the quantum photovoltaic yields, and pay little attention to the difference between the two photovoltaic processes in this DQDs photocell model.

\section{Results and discussions}

\begin{table}[htp]
\begin{center}
\caption{Parameters for this DQDs photocell.}
\setlength{\tabcolsep}{5mm}{
\begin{tabular}{cccc}
\hline& \(Parameters\)        & \( Units \) & \(Values\)  \\
\hline& $\mu_{\alpha\beta}$   &  eV         & 1.4         \\
        & $\omega$            &  eV         & 1.64        \\
        & $\varepsilon$       &  eV         & 0.012       \\
        & $\Gamma_i$          & eV          & 0.12        \\
         & $\Gamma_{ic}$      & eV          & 0.024       \\
         & $\gamma_{ic}$      & eV          & 0.02        \\
         & $\gamma_{1h}$      & eV          & $6.2*10^{-7}$    \\
         & $\gamma_{2h}$      & eV          & $1.98*10^{-7}$   \\
         & $n_{ic}$           &            & $4.57*10^{-4}$    \\
         & $N_{ic}$           &            & $4.57*10^{-4}$    \\
         & $N_{1h}$           &            & $1*10^4$          \\
         & $N_{2h}$           &            & $6*10^4$          \\
\hline
\label{Table1}
\end{tabular}}
\end{center}
\end{table}

\begin{figure}[htp]
\center
\hspace{0in}%
\includegraphics[width=0.35\columnwidth]{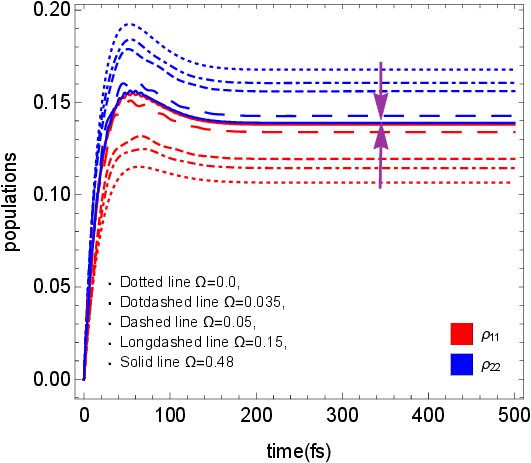}
\caption{(Color online) The time evolution of the photo-generated carriers populations \(\rho_{11}\)(Red curves), \(\rho_{22}\) (Blue curves) regulated by different tunneling coefficients $\Omega$ at room temperature $T_a$=300 K. Other parameters are taken from Table \ref{Table1}.}
\label{Fig.2}
\end{figure}

Owing to the aim to the tunneling effect on the quantum yields, we firstly focus on the photo-generated carriers transport in this DQDs photocell. Thereupon, the populations on the excited states \(|1\rangle\) and \(|2\rangle\) are the objects of interest to us. The selected parameters listed in Table \ref{Table1} are motivated by Ref.\cite{2013Photosynthetic} for this photocell model. First, we investigate dependence of time evolution of the photo-generated carriers populations \(\rho_{11}\) (Red curves), \(\rho_{22}\) (Blue curves) on the tunneling coefficients $\Omega$ at room temperature $T_a$=300 K in Fig.\ref{Fig.2}.

The curves in Fig.\ref{Fig.2} show the characteristics with climbing to a peak quickly and gradually flattening out, which manifests a vivacious photoelectric conversion behavior in this DQDs photocell. Notably, the photo-generated carriers populations \(\rho_{11}\) are increasing with the tunneling coefficient $\Omega$, while the carriers populated on the excited state \(|2\rangle\) are decreasing with $\Omega$, as can be got by two arrows facing each other in Fig.\ref{Fig.2}. Specially, when the tunneling coefficient $\Omega$=0.48, the curve of \(\rho_{11}\)(Red curve) almost overlaps with \(\rho_{22}\) (Blue curve). What is the underlying physics for this phenomena? If we turn our attention to the parameters, such as $\gamma_{1h}$=$6.2*10^{-7}$ eV, $\gamma_{2h}$=$1.98*10^{-7}$ eV in Table \ref{Table1}, maybe we can find its answers.
The larger $\gamma_{2h}$ means more photons with lower-energy can be absorbed in the transition \(|2\rangle\)\(\leftrightarrow\)\(|b\rangle\), and the excited state \(|2\rangle\) will obtain a larger population. However, the phenomenon will be changed by the increasing tunneling effect via the carriers transporting to \(|1\rangle\), which brings out the repopulation of photo-generated carriers between two QDs. What's going to happen with the redistribution of population? That is a matter of concern in this DQDs photocell system at room temperature.

\begin{figure}[htp]
\center
\hspace{0in}%
\includegraphics[width=0.35\columnwidth]{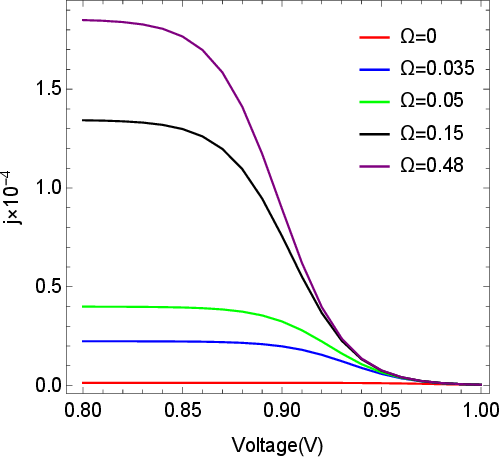}\includegraphics[width=0.3495\columnwidth]{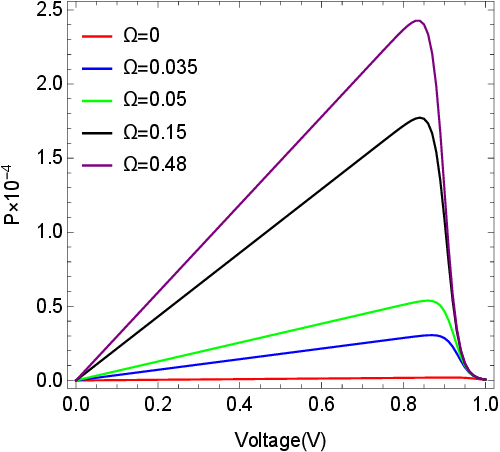}
\caption{(Color online) Current-voltage and power-voltage characteristics generated by the DQD photocell with different tunneling coefficients at room temperature $T_a$=300 K. Other parameters are the same to Fig.\ref{Fig.2}.}
\label{Fig.3}
\end{figure}

In the following, the current-voltage (j-V) and power-voltage (P-V) characteristics at $T_a$=300 K are shown in Fig.\ref{Fig.3}, respectively. The curves in Fig.\ref{Fig.3} respond to our concerns about the photovoltaic properties of this DQDs photocell. In the J-V characteristic curves, the areas under the curves reflect the output powers of this DQDs photocell system. It notes that the increasing output power and increasing short-circuit current are achieved by the increments of tunneling coefficients in the j-V characteristic curves. In the P-V characteristic illustrations of Fig.\ref{Fig.3}, the peak values of output power are increasing with the tunneling coefficients, but the open-circuit voltages are nearly identical shown by the P-V curves. Owing to the photovoltaic performances shown under the same parameter conditions in Fig.\ref{Fig.2} and Fig.\ref{Fig.3}, these results demonstrate that the repopulation of photo-generated carriers can indeed help to enhance the photovoltaic performances in this photocell system.

\begin{figure}[htp]
\center
\hspace{0in}%
\includegraphics[width=0.3\columnwidth]{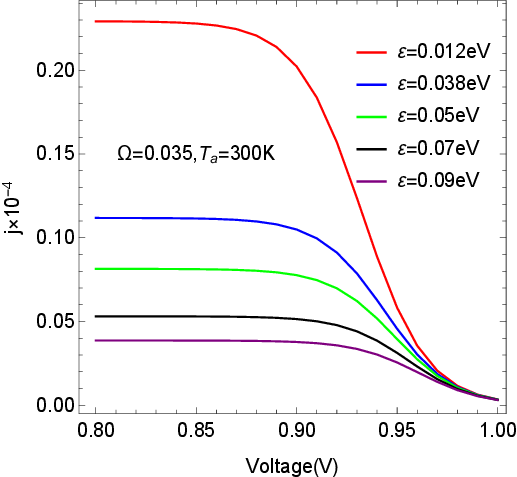}\includegraphics[width=0.3\columnwidth]{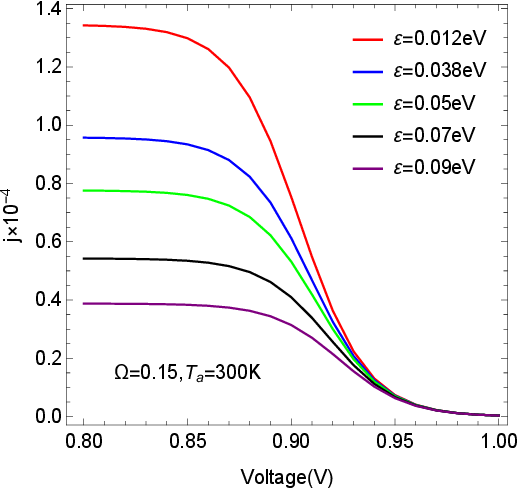}\includegraphics[width=0.29\columnwidth]{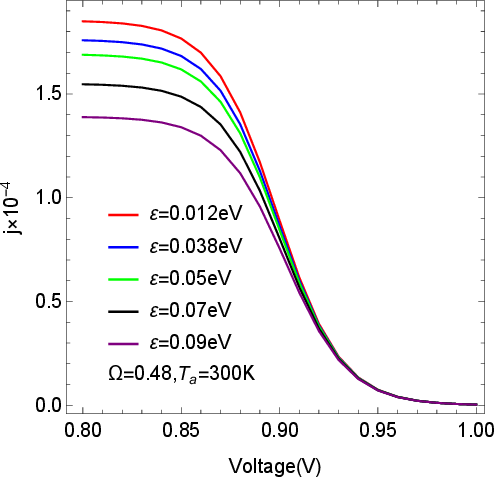}
\caption{(Color online) Current-voltage characteristics with different energy mismatch \(\varepsilon\) regulated by different tunneling coefficients $\Omega$ at room temperature $T_a$=300 K. Other parameters are taken from Table \ref{Table1}.}
\label{Fig.4}
\end{figure}

\begin{figure}[htp]
\center
\hspace{0in}%
\includegraphics[width=0.3\columnwidth]{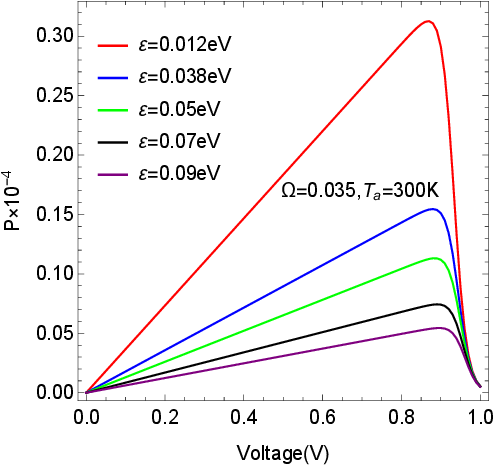}\includegraphics[width=0.3\columnwidth]{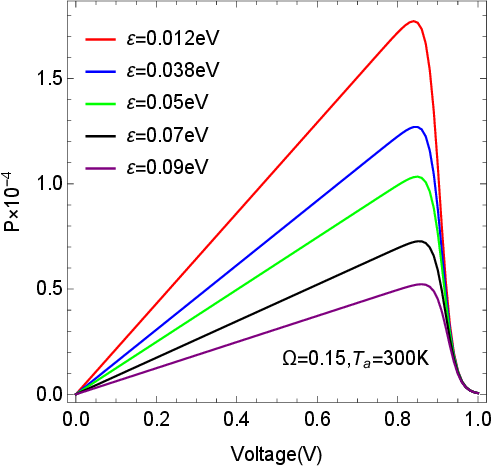}\includegraphics[width=0.3\columnwidth]{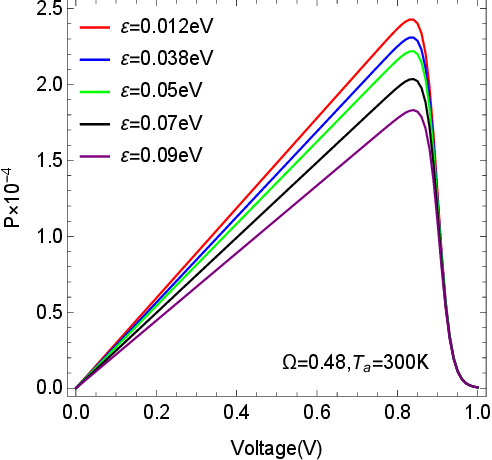}
\caption{(Color online) Power-voltage characteristics with different energy mismatch \(\varepsilon\) regulated by different tunneling coefficients $\Omega$ at room temperature $T_a$=300 K. Other parameters are the same to Fig.\ref{Fig.4}.}
\label{Fig.5}
\end{figure}

In fabricating a DQDs photocell, the energy mismatch \(\varepsilon\) between two QDs is a key parameter, and the role of \(\varepsilon\) in the photovoltaic characteristics attracts our attention. Therefore, the photovoltaic properties dependent on \(\varepsilon\) with different tunneling coefficients are discussed in Fig.\ref{Fig.4} and Fig.\ref{Fig.5}. The coordinate values on the vertical axis show the maximum short-circuit currents increasing with the tunneling coefficients \(\Omega\) tuned by 0.035, 0.15, 0.48 in Fig.\ref{Fig.4}.
However, the energy mismatch \(\varepsilon\) exhibits a negative impact on the photovoltaic performances. In Fig.\ref{Fig.4}, the nearly horizontal curves show that the maximum short-circuit currents are generated by the less energy mismatch \(\varepsilon\)=0.012eV with the same tunneling coefficient. What's more, the intervals between the maximum short-circuit currents, are gradually decreasing with the increments of \(\Omega\) by 0.035, 0.15, 0.48 in Fig.\ref{Fig.4}. It verify that the passive effect caused by \(\varepsilon\) can be reduced by the tunneling effect. These results are further proved to be true by the peak powers in the P-V characteristic in Fig.\ref{Fig.5}. Hence, we argue that the different roles between the tunneling coefficients \(\Omega\) and energy mismatch \(\varepsilon\) in photovoltaic properties originate from their different underlying physical regimes. We argue that the increasing radiation recombination rate(RRR)\cite{2020Radiative,2020Inhibited} of photo-generated carriers caused by the increasing energy mismatch \(\varepsilon\) is the main physical regime. Thereupon, the photovoltaic properties are weakened by the electron-hole recombination enhanced by the increasing energy mismatch \(\varepsilon\).

\begin{figure}[htp]
\center
\hspace{0in}%
\includegraphics[width=0.31\columnwidth]{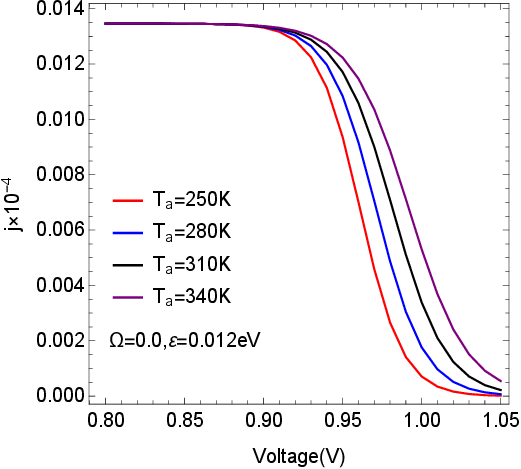}\includegraphics[width=0.3\columnwidth]{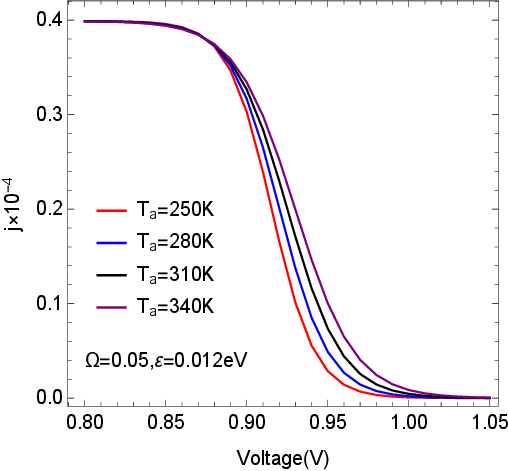}\includegraphics[width=0.3\columnwidth]{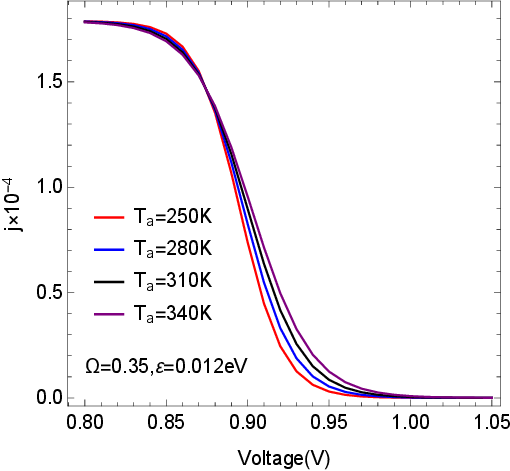}
\caption{(Color online) Current-voltage characteristics with different ambient temperatures $T_a$ tuned by different tunneling coefficients of $\Omega$, and other parameters are taken from Table \ref{Table1}.}
\label{Fig.6}
\end{figure}

\begin{figure}[htp]
\center
\hspace{0in}%
\includegraphics[width=0.31\columnwidth]{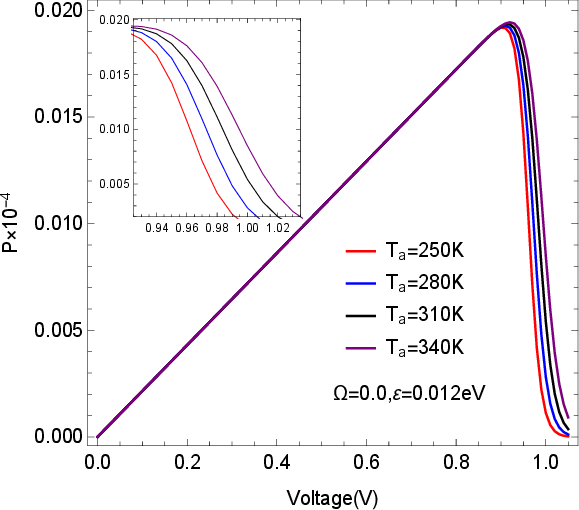}\includegraphics[width=0.3\columnwidth]{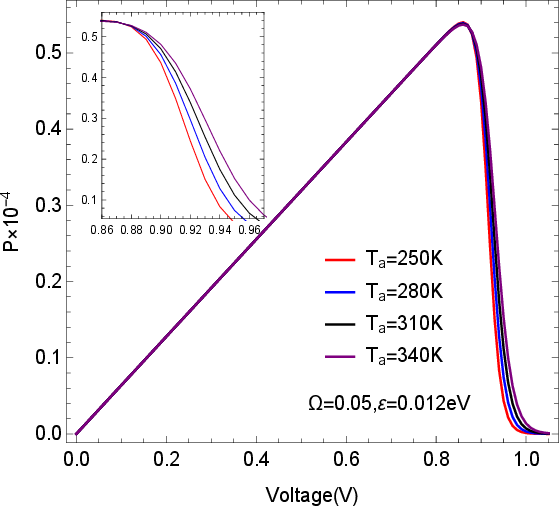}\includegraphics[width=0.3\columnwidth]{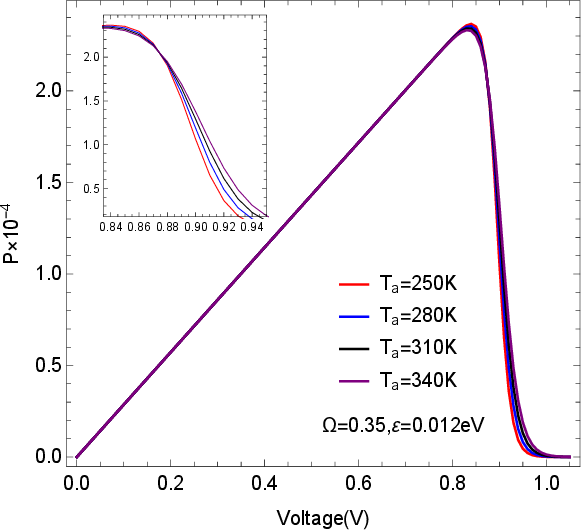}
\caption{(Color online) Power-voltage characteristics with different ambient temperatures $T_a$ tuned by different tunneling coefficients of $\Omega$, and other parameters are the same to Fig.\ref{Fig.6}.}
\label{Fig.7}
\end{figure}

The ambient environment temperature has been proved to be an important parameter for efficient QDs photovoltaic cells\cite{Zhao2019}.
In the following, Fig.\ref{Fig.6} and Fig.\ref{Fig.7} show the role of ambient temperature $T_a$ in the photovoltaic performance of the DQDs photocell. At $T_a$=300 K, in the process of the tunnel coefficient $\Omega$ increasing from 0.0, 0.05 to 0.35, the maximum short-circuit currents, output powers in Fig.\ref{Fig.6} and the peak output powers in Fig.\ref{Fig.7} increase with these increments, except for the gradual decrements in the differences between the peak powers in Fig.\ref{Fig.7}. Not only that, the ambient temperature $T_a$ also positively affects its photovoltaic performances. From Fig.\ref{Fig.6} and Fig.\ref{Fig.7}, we noticed that the output powers and open-circuit voltages increase with the increments of ambient temperature $T_a$, as can be drawn from the inner illustrations in Fig.\ref{Fig.7}. But at the same time we also noticed that $T_a$ has little effect on the maximum short-circuit currents in Fig.\ref{Fig.6} and peak powers in Fig.\ref{Fig.7} with the same tunneling coefficient. We tentatively put forward that the transport rate of carriers can be greatly improved by $T_a$, which brings about more carriers quickly accumulating on the external electrodes. Therefore, the increasing open-circuit voltages and output powers are generated. However, the enhancement of $T_a$ on the transporting of carriers, and the recombination rate of carriers eventually reach an equilibrium with the same tunneling coefficient $\Omega$. Hence, the continuous increments of $T_a$ cannot change the maximum short-circuit currents in Fig.\ref{Fig.6} and peak powers in Fig.\ref{Fig.7}.

\section{Conclusion}

To summarize, in this work we investigated the tunneling effect on quantum yields of a coupled QD pairs. We explicitly identified the photovoltaic properties, such as the short-circuit current, open-circuit voltage and output power dependent on the tunneling coefficient, energy mismatch and the ambient environment temperature via evaluating the current-voltage and power-voltage characteristics. The most significant result is the influence of tunneling effect on the photo-generated carriers populations in the two excited states. The increasing tunneling effect leads to the repopulation in the double excited states at the room temperature, which brings out the increasing output power, short-circuit current and peak power.
In addition, the quantum yields can be enhanced by the increasing ambient temperatures but shrunk by the increasing energy mismatches between two QDs. However, the passive influence cause by the energy mismatches can be greatly reduced by the tunneling effect.

These results not only provide insights into the photovoltaic performances of the DQDs photocell system, but also provide some artificial strategies for efficient assembled QD arrays photocell via the tunneling effect inspired by this DQDs photocell. Specially, Within current experimental development, our results about the regulated scheme may probably be implemented by the external bias voltage\cite{Zhao2009Manipulative}, which can provide some significant practical suggestions to enhance quantum yields in other artificial photovoltaic devices.

\section*{Acknowledgments}

\noindent S.-C. Zhao is grateful for funding from the National Natural Science Foundation of China (Grants 62065009 and 61565008).

\section*{Conflict of interest}
\noindent The authors declare that they have no conflict of interest.

\bibliography{reference}
\bibliographystyle{unsrt}
\end{document}